# Phase transitions in layered crystals

Yuri Mnyukh
*76 Peggy Lane, Farmington, CT, USA, e-mail: yuri@mnyukh.com*
(Dated: May 21, 2011)

It is demonstrated by analyzing real examples that phase transitions in layered crystals occur like all other solid-state phase transitions by nucleation and crystal growth, but have a specific morphology. There the nucleation is *epitaxial*, resulting in the rigorous orientation relationship between the polymorphs, such that the directions of molecular layers are preserved. The detailed molecular mechanism of these phase transitions and formation of the laminar domain structures are described and related to the nature of ferroelectrics.

## 1. Layered structures

Specifics of phase transitions in layered crystals will be demonstrated by analyzing two examples: *hexamethyl benzene* (HMB) and *DL-norleucine* (DL-N). They differ in their properties, molecular shape, and crystal structure.

HMB, $C_6(CH_3)_6$, is an aromatic substance with flat circular "coin-like" molecules (Fig. 1a). This molecular shape allowed a very energetically advantageous close packing into pseudo-hexagonal molecular layers, a molecular plane coinciding with the layer plane [1,2]. All intermolecular bonding, both in the layers and between the layers, is of a Van der Waals' type. The minimum distance between the carbon atoms of the benzene rings in adjacent (*001*) molecular layers is larger than the sum 3.40Å of carbon Van der Waals' radii due to repulsion of the $CH_3$ groups. As a result, the interaction between the layers is weakened, giving rise to the layered structure (Fig. 1b).

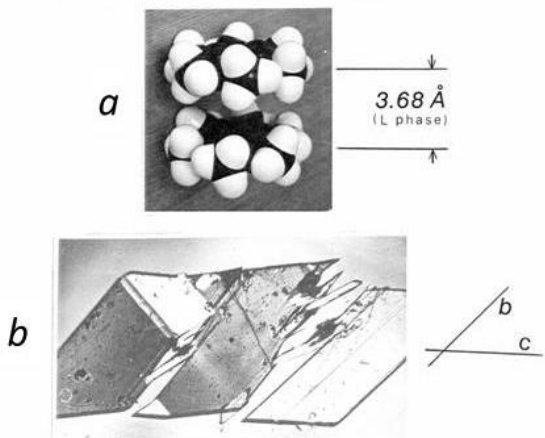

Fig. 1. Layered structure of hexamethyl benzene (HMB) crystals.
(a) HMB molecules in adjacent (*001*) layers, illustrating why HMB has layered structure: the molecular shape prevents sufficiently close interlayer packing.
(b) Lamination of a crystal along (*001*) when it is pricked with a needle.

DL-N is a short-chain aliphatic substance, $CH_3 \cdot (CH_2)_3 \cdot CHNH_2 \cdot COOH$, with a layered crystal structure typical of chain molecules, where the molecular axes are quite or almost perpendicular to the layer plane [3]. Each layer is bimolecular: the CNCOO groups of the molecules are pointed toward the center of the layer where they form a network of hydrogen bonds N-H... O [4] (Fig. 2). This central "skeleton" turns the bimolecular layer into a rather firm structural unit. The interlayer interaction is comparatively weaker, because it is of a purely Van der Waals' type, so the layer stacking is governed exclusively by the principle of close packing. As a result, both DL-N polymorphs have a pronounced layered structure of almost the same layers in different stacking.

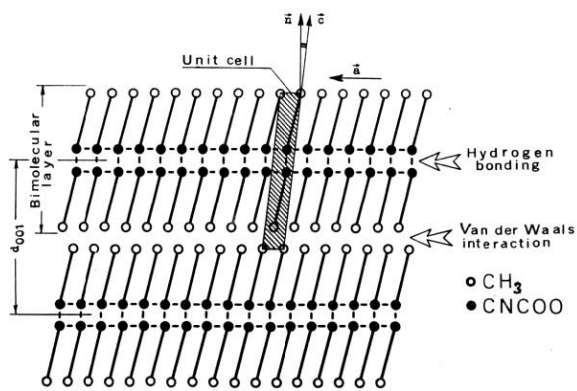

Fig.2. Characteristic features of the DL-norleucine (DL-N) crystal structure [4]. The layer spacing in the lower-temperature phase is $d_{001} = 16.03$ Å.

In general, a layered structure has strongly bounded, energetically advantageous two-dimensional units − molecular layers. Since the layer stacking contributes relatively little to the total lattice energy, the difference in the total free energies of the structural variants is small. This is a prerequisite for the polymorphism in layered crystals. Change from one polymorph to the other is mainly reduced to the mode of layer stacking.

The layers themselves only slightly modified under the influence of different layer stacking.

## 2. Nucleation-and-growth phase transitions

Prior to dealing with phase transitions in layered crystals, the molecular mechanism of phase transitions.in non-layered crystals needs to be outlined. It has been revealed in the studies [5-17] summarized in [18]. It is a *crystal growth* involving nucleation and propagation of interfaces (Fig. 3), very much similar to crystal growth from liquids. Molecular model of the interface and the molecule-by-molecule structural rearrangement leading to its propagation is shown in Fig. 4. The main feature of this mechanism is *a sufficient steric freedom of the molecular rearrangement still in the gravitation field of the new phase* [12].

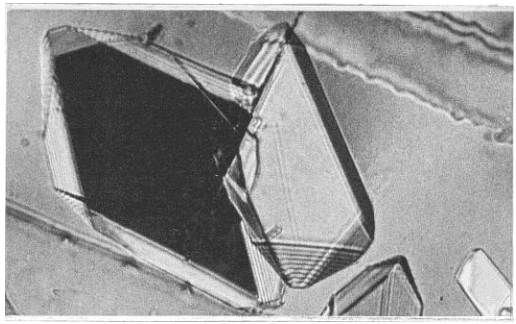

Fig.3. Phase transition from low-to-high temperature phase (L→H) in *p*-dichlorobenzene (PDB). Four separate H single crystals of different orientations are growing within the L single crystal (background). Two largest ones have grown into one another as a result of competing for the building material that the surrounding L crystal is. Absence of orientation relationship between H crystals, as well as between them and L, is obvious. Evidently, the H *nuclei* had been oriented arbitrarily in the L lattice.

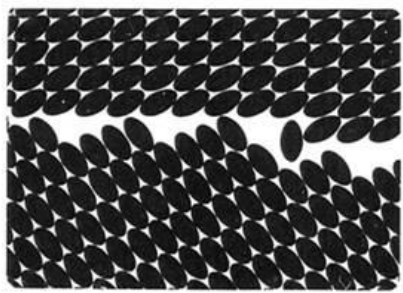

Fig. 4. Molecular model of phase transition in a crystal. The *contact* interface is a rational crystal plane in the resultant phase, but not necessarily un the initial phase. The interface advancement has the *edgewise* mechanism: it proceeds by shuttle-like strokes of small steps (kinks), filled molecule-by-molecule, and then layer-by-layer in this manner. (Crystal growth from liquids is realized by the same mechanism). Besides the direct contact of the two different structures, existence of the 0.5 molecular layer gap on average should be noted. It is wide enough to provide steric freedom for the molecular relocation (only at the kink), but it is narrow enough for the relocation to occur under attraction from the resultant crystal. More detailed description of the process and its advantages is given in Ref. 18 (Sec. 2.4.2 - 2.4.6 ).

The same principle is applied to the nucleation. It is not the classical fluctuation-based nucleation described in textbooks. It is pre-determined. The nuclei are located at specific crystal defects - microcavities of a certain optimum size. The microcavities provide sufficient steric freedom for the molecular relocation and, at the same time, assistance to that relocation by molecular attraction from the opposite side of the cavities. The distinctions in the size and shape of the microcavities determine both the individual nucleation temperature $T_n$ of each nucleation site and the crystal orientation of the new phase growing from it.

## 3. Phase transition in HMB
.
It had been reported [19] that HMB phase transition L→H at about 110 $^oC$ occurs "instantaneously" without changing the direction of light extinction when it was observed under crossed polarizers. The more detailed investigation of the transition with crystals of good quality [14,18] revealed the following. First, a more precise temperature $T_o$ (110.8 $^oC$) corresponding to equal free energies of the polymorphs was established. Then it was found that the transition occurs not instantaneously, but by nucleation and gradual growth of the H phase. The nucleation occurred at crystal defects and exhibited hysteresis. The lags $\Delta T_{tr} = T_{tr} - T_o$ were greater in more perfect crystals. Nucleation could be initiated at any point by a touch with a needle. The interfaces were observable (Fig. 5). Their movement could be halted by lowering $\Delta T_{tr}$ to zero. Even small increases in $\Delta T_{tr}$ sharply accelerated interface motion (Fig. 6). At $\Delta T_{tr} > 2.5$ $^oC$ the interfaces advanced at the rate > 2 mm/sec. With nucleation lags of that order or higher (a realistic assumption), the phase transition in single crystals or grains of 0.4 mm size would be completed within 0.2 sec and appear instantaneous. Such an "instantaneous" transition is still $10^5$ -$10^6$ times slower than the velocity of elastic wave in a solid medium. The first transition was initiated at several points on the edges, continued by formation of thin H strips parallel to the cleavage, and then proceeded by a gradual width increase of the H bands denoted by shading between the arrows. The frontal advancement of the interfaces visible on the photographs was not, however, truly gradual; it rather proceeded by lateral strokes of very small steps along the interface lines (*edgewise* mechanism, like in Fig. 4), as a closer visual examination revealed.

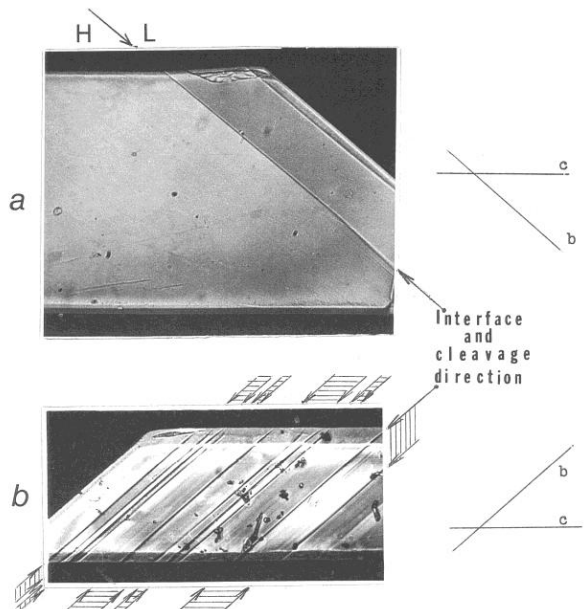

Fig.5. The interfaces (arrows) during phase transitions in HMB crystals. Crossed polarizers. Upon rotation of the microscope stage both phases were extinguished simultaneously.
(a) The first transition (L → H) in a fairly perfect single crystal. The interface remained parallel to itself and the cleavage planes.
(b)The last of the cyclic phase transitions
(L → H → L → H → ) L → H.

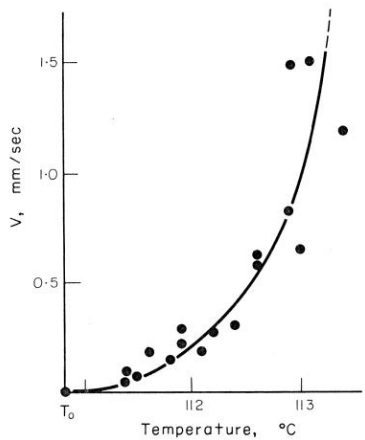

Fig. 6. The temperature dependence of velocity V of interface motion in HMB phase transition. $T_o = 110.8\ ^oC$.

There can be no doubt in the nucleation-and-growth mechanism of the HMB phase transition. But its morphology (Fig. 5) differs from that in Fig. 3. This time the interfaces have always the same direction parallel to the molecular layers and cleavage.

Nucleation occurs at the layer edges and initially gives rise to formation of narrow wedges of the H phase with numerous growth steps along a wedge generating line (Fig. 7a). The wedges then penetrate through the crystal to form bands of the H phase seen in Fig. 5b. An important feature of this growth morphology is the *edgewise* mechanism as illustrated in Fig. 4. The molecular layers do not slip as a whole over one another.
Rather, *every layer is subjected to a complete rearrangement* (Fig. 7b). The x-ray Laue patterns confirmed a strict orientation relationship of the phases, while the optical examination with crossed polarizers made it clear that the layer orientation has not changed (Fig. 7c).

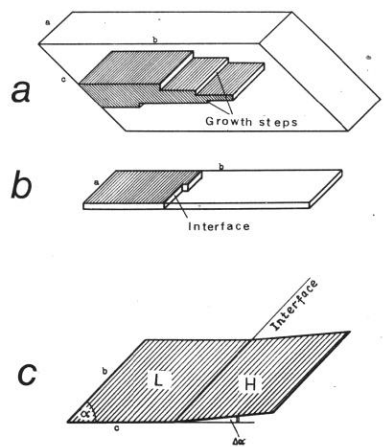

Fig. 7. Morphology of the HMB phase transition (schematic). See text.

### 4. Phase transition in DL-N

After discovery of the DL-N phase transition at ~117 $^oC$ [14] the first impression was that it may occur without hysteresis. In such a case it would not represent nucleation and growth. The task was to verify whether at least one of the sufficient indicators of the nucleation-and-growth mechanism is present. These indicators are [20] hysteresis, phase coexistence, interface motion, ability to initiate the transition by mechanical disturbance (note: in fact, these are different forms of one and the same indicator). To this end, smaller and more perfect single crystals were prepared. They were 0.5 to 2 mm size rhombus- or trapezium-shaped plates as thin as 0.02 to 0.1 mm. With these crystals and temperature control better than $0.1 ^oC$ the nucleation hysteresis was detected, although it was rather small. Fig. 8 attests that all L → H transitions start at $T > T_o$, while H → L at $T < T_o$. Initially $|\Delta T|$ is about 0.8 $^oC$, then decreases to stay at 0.2 $^oC$ level. In another experiment, $\Delta T_n$ was compared in pairs of one visually perfect and one less perfect single crystal of equal size. For each of the 10 pairs examined there was $|\Delta T_n|_{perf} > |\Delta T_n|_{imperf}$ by 0.3 to 0.8 °C. Finally, introduction of a mechanical defect initiated transition at a lower ΔT. All the observations indicated nucleation at the crystal imperfections.

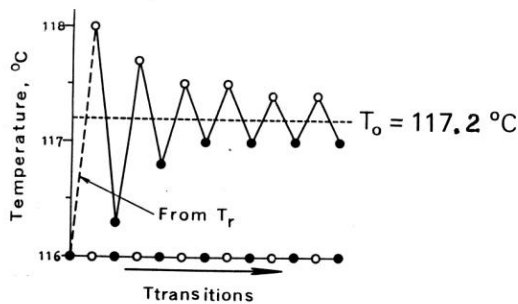

Fig. 8. Temperature hysteresis ΔT upon cyclic phase transitions in a DL-N crystal. The hysteresis is small and decreases with the number of the transitions to a low, but not zero, level.

When a (*001*) face was viewed in microscope under crossed polarizes, the direction of maximum extinction of the crystal was preserved after phase transition, thus showing a rigorous structural orientation relationship between the phases. Laue photographs taken before and after phase transition were almost indistinguishable, which proves both the rigorous orientation relationship and a strong similarity between the two structures. Then a set of the powder photographs as a function of temperature was taken and used to find the intralayer spacings $d_{200}$ and $d_{020}$ *vs.* temperature. There was a minute but noticeable (~1%) change in these spacings, which meant that the layers in the two phases are not completely identical.

The morphology of the phase transition was much like that in HMB. In an imperfect crystal, being actually a stack of weekly bound lamellae parallel to the molecular layers, the linear interfaces of one and the same direction moved separately and poorly coordinated in different lamellae. But in the relatively perfect crystals consisting of only few lamellae the linear interfaces were sharp. Their motion could be controlled by carefully regulated temperature. During this process every lamella was divided by the moving interface in two phases.

Finally, the long spacing $d_{001}$, which is a characteristic of the layer stacking, was observed on a screen and photographed upon heating over the transition range (Fig. 9). *Two distinct phases of different layer stacking coexisted throughout a range of transition.* The L phase was represented by the interlayer spacing $d_{001}$=16.5Å. Upon heating, the second line representing the interlayer spacing $d_{001}$ =17.2 Å of the H phase appeared. Its intensity gradually increased from zero to a maximum at the expense of the intensity of the L line until the latter was completely extinguished. This experiment decidedly refuted any idea of a gradual qualitative change.

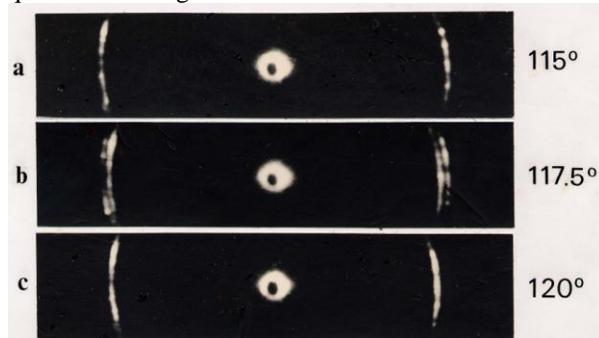

Fig. 9 Change in the interlayer spacing $d_{001}$ upon the L → H phase transition in a DL-N powder specimen. The photographs were taken from the screen of a device for direct viewing small-angle X-ray patterns.
(a) Before the transition. $d_{001}$ = 16.5 Å.
(b) During the transition. Two separate lines coexist, each representing one of the phases. It was visually observed that the intensity of the H line was gradually increasing at the expense of the L line. Thus, the H *quantity* in the heterophase specimen was increasing, and L decreasing over the temperature range. This experiment visualizes how the apparent "continuous" and "displacive" phase transitions really proceed.
(c) After the transition; $d_{001}$ has increased by 4.1%.

## 5. Epitaxial nucleation in interlayer microcracks

HMB and DL-N differ in the molecular shape and chemistry, but exhibit similar features of their phase transitions not found in those described in Section 2. *This is due to their layered crystal structure.* These transitions occur by nucleation and growth as well. Nucleation requires presence of optimum-sized microcavities  In layered structures the interlayer interaction is weak on definition. In practice, layered structures always have numerous defects of imprecise layer stacking. Most of these defects are minute wedge-like interlayer cracks on the crystal faces as viewed from the side of layer edges. In such a microcavity there always is a point where the gap has the optimum width for nucleation. There the molecular relocation from one wall to the other occurs with no steric hindrance and, at the same time, with the aid of attraction from the opposite wall. In view of the close structural similarity of the layers in the two polymorphs, *this nucleation is epitaxial.* In accordance with this description, the nucleation was indeed seen initiated at those crystal faces (Fig. 5 and 7). Fig. 10 is a schematic of initial stage of the epitaxial phase transition. Orienting effect by the substrate ensures preservation of the direction of molecular layers.

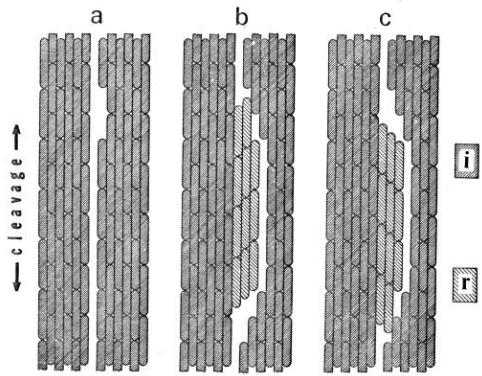

Fig. 10. Initial stage of epitaxial phase transition (schematic).
(a) A lattice defect in the form of a submicroscopic crack parallel to the cleavage.
(b) Oriented embryo of the resultant phase, formed by consecutive transfer of molecules from one side of the microcavity to the other.
(c) Equally probable embryo in the "twin" orientation relative to its counterpart; it can form if the resultant lattice has a lower symmetry.

Finally, there is a simple answer to why the hysteresis $\Delta T_n$ in epitaxial phase transitions is small (Fig. 8) as compared with non-epitaxial transitions. Due to the abundance of wedge-like microcracks, there is no shortage in the nucleation sites of optimum size; at that, the presence of a substrate of almost identical surface structure acts like a crystallization "seed". Therefore only small overheating or overcooling is required in order to initiate phase transition. Without a scrupulous experimental verification, the phase transitions in question may be taken for "displacive", "instantaneous", "cooperative", "soft-mode", "second-order", etc.

## 6. Displacive phase transition by nucleation and growth

The idea of *displacive* phase transitions was put forward in 1950's as a cooperative displacement of atoms/molecules in the crystal lattice without breaking their bonding. At that, a rigorous structural orientation relationship was assumed without saying. The alternative was "reconstructive" phase transitions if such structural modification could not be imagined without breaking bonds; how the latter can occur remained unknown. The classification was not based on experimental investigations of the process. It was simply assumed from comparisons of the initial and final structures. The comparisons frequently resulted in "hybrid" cases with some bonds being broken; those cases were deemed "displacive" anyway. The adjective "displacive" is now loosely applied to the cases where the structures of polymorphs seem to be "sufficiently similar".

In terms of a structural comparison of their polymorphs, the phase transitions in HMB and DL-N are the most suitable candidates to be *displacive* by displacing the molecular layers over one another to the new layer stacking. Indeed, a strict orientation relationship there is an experimental fact, and the bonding inside the layers remains unchanged. But, speaking figuratively, nature does not take advantage of making their phase transitions *displacive*. The transitions are a crystal growth on every account: nucleation, moving interfaces, phase coexistence in a temperature range, hysteresis, and discontinuous change (2.4% and 6%) in the specific volumes. Every molecular layer undergoes a reconstruction, molecule by molecule, to build up a new layer of almost the same structure, but now in a different layer stacking. Even the cementing action of hydrogen bonding inside the DL-N molecular layers does not prevent them from that reconstruction. The observational indicators of this kind of crystal growth is (a) rigorous orientation relationship of the phases, (b) a uniform direction of the interfaces, and (c) a relatively low hysteresis $\Delta T_n$. These features can be understood in terms of the *nucleation-and-growth* mechanism combined with the *nucleation epitaxy*. It is the most energy-efficient mechanism, considering that it needs energy to relocate only one molecule at a time, and not the myriads of molecules at a time as any cooperative change requires.

## 7. Laminar domain structures

It had been noticed [21-23] that phase transitions in layered crystals produce L phase in different equivalent orientations, approximately in equal quantities. The phenomenon was interpreted as the consequence of a *displacive* mechanism acting in different directions in different parts of the crystal. However, these phase transitions occur by *epitaxial crystal growth*. Formation of the laminar-domain structures is almost inevitable in epitaxial transitions if the emerging phase has a lower symmetry. The phenomenon is observed only in H → L transitions because it is L that has a lower symmetry. As illustrated in Fig. 10, an oriented L nucleus can appear with equal probability in two orientations related to one another as crystallographic twins. Fig. 11 shows this in more detail.

Growth of a nucleus gives rise to formation of a *laminar domain* of one or the other orientation. As a rule, phase transitions in layered crystals are multinuclear. Approximately one half of the laminar domains assumes the orientation No.1, and the rest assumes No.2. When the adjacent domains of the same orientation meet, they merge into a single domain. The laminar-domain structure with the strict alternation of No.1 and No.2 orientations, sketched in Fig.11b, emerges. Different domains are not uniform in thickness, but any two adjacent domains are related as crystallographic twins.

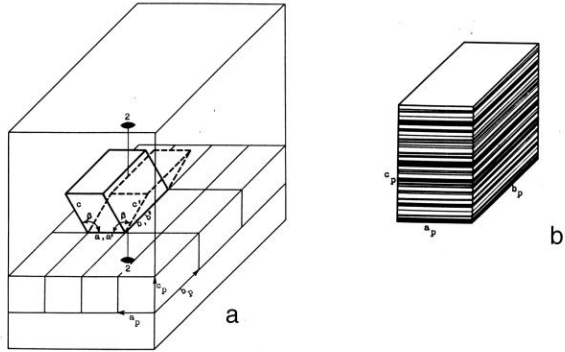

Fig. 11. Formation of "twin" domain structure in epitaxial phase transitions as a result of multinucleation of a lower-symmetry phase. The nuclei assume two equally probable orientations.
(a) Two equally probable orientations of a monoclinic lattice of the resultant phase in the initial crystal characterized by a rectangular lattice and cleavage (*001*). They can be brought into coincidence with a two-fold axis perpendicular to (*001*).
(b) Growth of each nucleus leads to formation of a lamina in one of the two possible orientations. The initial single crystal turns into a laminar structure of the domains of two alternating orientations shown as black and white.

## 8. Why ferroelectrics are not pyroelectrics

Pyroelectrics and ferroelectrics are both spontaneously polarized dielectrics, but only the latter have the ability to be polarized / repolarized by the applied electric fields. The difference can now be explained: only ferroelectrics have a layer structure. Nucleation in *ferroelectric↔paraelectric* phase transitions and in rearrangements of the laminar-domain systems is *epitaxial*. The *epitaxial* nucleation has a sufficiently low activation energy to be controlled by the applied electric field. The resultant structural change brings about a new state of polarization.